\documentclass[a4paper,12pt,onecolumn]{article}

\usepackage[english]{babel}
\usepackage{fancyhdr} 
\usepackage[pdftex]{graphicx} 
\usepackage{url} 
\usepackage{caption} 
\usepackage{subcaption} 
 \usepackage{natbib} 
\usepackage{multirow} 
\usepackage{float} 
\usepackage{amssymb}
\usepackage[intlimits]{amsmath} 
\usepackage{amsthm} 
\usepackage[german]{nomencl}
\usepackage{booktabs} 
\usepackage{geometry}
\usepackage{enumitem}
\usepackage{bbm}
\usepackage{array}
\usepackage{authblk}

\makenomenclature
\usepackage[pdfborder={000},colorlinks=true,linkcolor=blue,citecolor=blue]{hyperref}
\usepackage{listings}
\usepackage[table]{xcolor}
\usepackage{tikz}
\usetikzlibrary{shapes,arrows,shadows, positioning}
\usepackage{bm,times}
\usepackage{smartdiagram}

\usepackage{attachfile}

\definecolor{b}{rgb}{0,0,.8}	
\definecolor{g}{rgb}{0,.6,0}	
\definecolor{n}{rgb}{0,0,0}	
\definecolor{h}{rgb}{0.4,0.2,0.2}	
\definecolor{v}{rgb}{0.2,0.6,0}

\newcommand{\bsI}{\boldsymbol I}

\newcommand{\bsX}{\boldsymbol X}
\newcommand{\bsY}{\boldsymbol Y}

\newcommand{\bseps}{\boldsymbol \varepsilon}

\newcommand{\bsGamma}{\boldsymbol \Gamma}
\newcommand{\bsPhi}{\boldsymbol \Phi}
\newcommand{\bsPsi}{\boldsymbol \Psi}

\newcommand{\bsPi}{\boldsymbol \Pi}

\newcommand{\ov}\overline
\newcommand{\what}{\widehat}

\newcommand{\rig}\right
\newcommand{\lef}\left
\newcommand{\nf}\normalfont

\definecolor{dcyan}{rgb}{0,0.5,.5}

\definecolor{dgreen}{rgb}{0,0.7,0}
 
\definecolor{dgrey}{rgb}{0.6,0.6,.6}

\title{On cointegration for modeling and forecasting wind power production
}  

\author{
	Florian Ziel$^{1}$\,,
	Antonia Arsova$^{2;§}$
}

\affil{
	\textit{$^{1}$University of Duisburg-Essen}\\
	\textit{$^{2}$TU Dortmund University}
}

\begin{document}
	\maketitle
\lhead{\nouppercase{\leftmark}}
\begin{abstract}

This study evaluates the performance of cointegrated vector autoregressive (VAR) models for very short- and short-term wind power forecasting. Preliminary results for a German data set comprising six wind power production time series indicate that taking into account potential cointegrating relations between the individual  series can improve forecasts at short-term time horizons. 

\end{abstract}


\section{Introduction and Motivation} \label{Introduction}
Wind power is becoming ever more important for the global energy supply owing to its carbon emissions-free electricity production capacity. Nevertheless, its dependence on meteorological conditions and inherent uncertainty pose risks to electrical grid operations and participants at electricity markets alike, thus calling for improved forecasts at various time horizons. 
In short-term wind power forecasting high-dimensional vector autoregressive (VAR) models are seen as competitive alternatives to other statistical methods, including machine learning techniques, see e.g. \cite{dowell2015very,ziel2016forecasting,browell2018improved} and \cite{messner2019online}. 
As these models have relatively low computational cost in comparison to more sophisticated non-linear models, they also serve as a suitable input for forecast combination methods and large scale problems.

The so-called 
\textit{persistent forecaster} (also \textit{naive forecaster}), which predicts the wind power output in the next time step with the last observed value, is used in many forecasting studies as the simplest benchmark \citep{ambach2016space, nazare2020wind}.
Despite its simplicity, it performs remarkably well and can thus also serve as a suitable candidate for forecast combination methods \cite{liang2016short, chen2017combined}. From a data scientist's point of view, the persistent forecaster corresponds to the multivariate model $\bsY_t = \bsY_{t-1} + \bseps_t$, which is known as a multivariate ($d$-dimensional) random walk. It can be regarded as a special case of the multivariate ARIMA process, which is an integrated process of order one ($\text{I}(1)$). When a linear combination of two or more integrated processes forms a stationary, or $I(0)$, equilibrium relationship between them, then these processes are known to be cointegrated. The number of linearly independent cointegrating relations between the $d$ integrated components is given by the \textit{cointegrating rank}. Thus cointegrating rank $0$  corresponds to the $d$-dimensional random walk, while cointegrating rank $d$ implies stationarity of all $d$ components of the process.
Linking the co-movements of non-stationary time series, cointegrating relations provide potentially useful information for modeling and forecasting, which can be utilized when the VAR model is cast in the form of a vector error correction model (VECM). 
As the well-performing\footnote{See, e.g., \cite{dowell2015very}.} stationary $\text{VAR}$-type model has a maximal cointegrating rank $d$, while the naive forecaster of minimal cointegrating rank $0$ performs satisfactorily as well, it makes sense to consider also models 'in-between', that is, VECMs with cointegrating rank $r=1, \ldots, d-1$. As they can be written as $\text{VAR}(p)$ models, it is reasonable to expect that they exhibit similar, if not better predictive performance than that of a VAR. Despite cointegration describing long-run phenomena, cointegrated VAR models and variants thereof have been useful also for short-term econometric forecasts, e.g. of spot and futures prices as in \cite{dolatabadi2018economic}.

\begin{figure}[h]
\includegraphics[width=.99\textwidth]{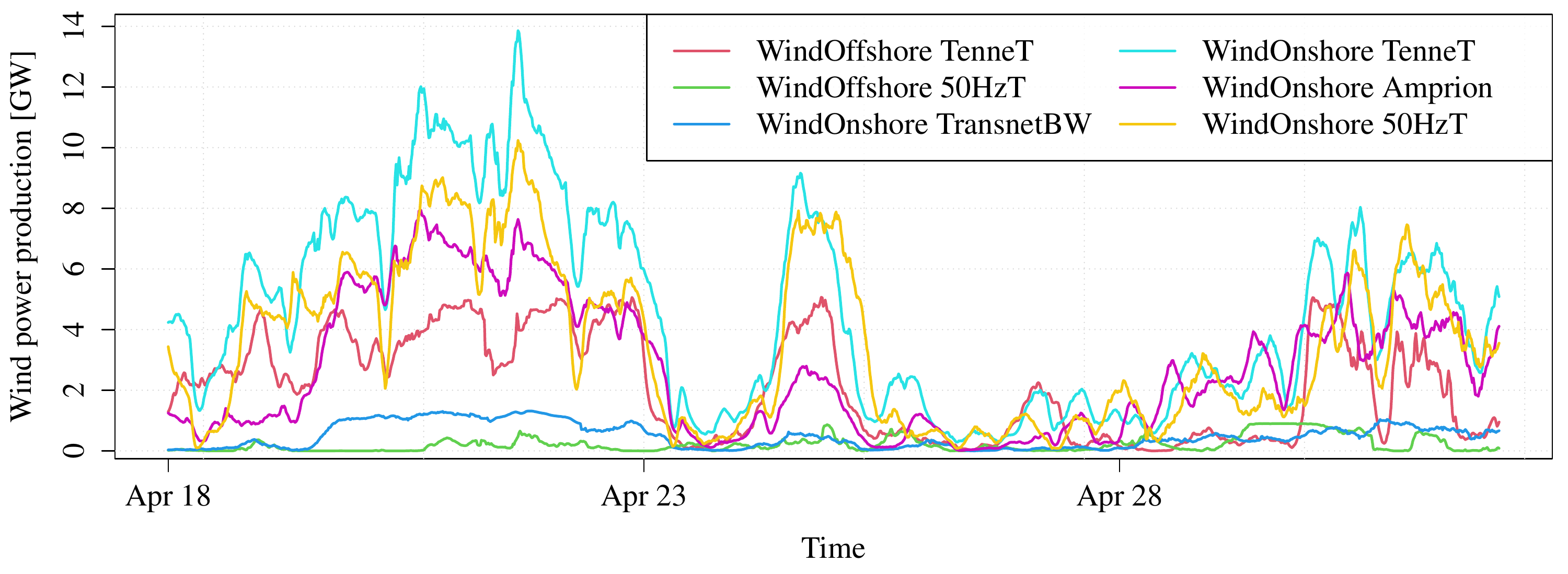} 
        \caption{Considered German wind power data for two selected weeks in 2020.}
        \label{fig_weights_1}
\end{figure}

It is widely accepted that wind power time series are non-stationary. This is mainly due to potential daily and annual seasonal pattern influenced by meteorological features \citep{ziel2016forecasting, ambach2016space}, but also to the aging of turbines, and on a regional level to the increase of installed wind power capacity.  However, on a local level (in the time domain) wind power time series appear to be close to a $I(1)$ process (Figure \ref{fig_weights_1}).


Despite having been a principal tool in econometric analysis for already more than three decades, the concept of cointegration has remained largely unexploited in the wind power forecasting literature. One current area of application is monitoring and failure detection at a turbine level, see, e.g. \cite{zolna2015nonlinear, dao2018operational, dao2018condition, zhang2020realization}. Another area is employing testing for cointegration in feature selection for machine learning models as in \cite{feng2017}.  However,
potential improvements in wind power forecasts owing to directly considering cointegrated VAR models have not been investigated so far. 

In this manuscript we aim to explore the integration and cointegration properties of wind power data and their potential for prediction models. First we briefly recap conintegration in VECMs. We then introduce the data set and explain the design of the forecasting study. Next we discuss the prediction results with focus on the effects of cointegration. We close with a short conclusion, outlining directions for future work.


\section{Model}

We consider a VECM for forecasting the $d$-dimensional vector of wind power production series $\bsY_t$. 
Every VECM can be written as in a VAR representation. 
A VAR($p$) model with external regressors  $\bsX_t$is given by
\begin{equation}
 \bsY_t = \bsPsi\bsX_t + \sum_{k=1}^p \bsPhi_k \bsY_{t-k} + \bseps_t.
\end{equation}
Using $\bsY_t = \bsY_{t-1} + \Delta \bsY_t $ with $\Delta$ denoting the first-difference operator, a cointegrated VAR model can be written in VECM representation as
\begin{equation}
\Delta \bsY_t = \bsPsi\bsX_t  + \bsPi \bsY_{t-1} + \sum_{k=1}^{p-1} \bsGamma_k \Delta \bsY_t + \bseps_t ,
\end{equation}
where $\bsPi = - \bsI + \sum_{k=1}^p \bsPhi_k $ and 
$\bsGamma_k = - \sum_{j=k+1}^p \bsPhi_k$ for $k\in\{1,\ldots,p-1\}$.
The rank $r=\text{rank}(\bsPi)$ corresponds to cointegrating rank of the VECM model. In the special case of a full rank $r=d$ the process $\bsY_t$ follows a VAR($p$) process, while for $r=0$ the difference process $\Delta \bsY_t$ follows a VAR($p-1$). The cases $0<r<d$ correspond to the standard cointegration cases.
We estimate the VECM using Johansen method, which corresponds to maximum likelihood estimation under the normality assumption for $\bseps_t$.

\section{Data and Forecasting Study}

We consider actual quarter-hourly German wind power production data from \textit{transparency.entsoe.eu}. The data set contains 
data for $d=6$ different wind power production regions categorized by the TSOs, including two off-shore regions and four on-shore regions. It covers a time span of about 4.5 years, from 1 January 2015 to 29 June 2020.

We conduct a rolling window forecasting study for the VECM models, with autoregressive order $p\in\{1,\ldots, 7\}$, cointegrating rank $r\in \{0,\ldots, d\}$ and 
calibration window length (in-sample size) $T\in\{96, 192, 384, 768, 1536, 3072\}$. The calibration window size corresponds to a sample size comprising 1, 2, 4, 8, 16 and 32 days. We perform $H$-step ahead forecasts with $H=8$, which is a forecasting horizon of two hours. To limit computational costs, we sample $N=1000$ randomly selected time points in the available data set which correspond to the assumed last known observation.

We evaluate the predictive performance using the multivariate mean absolute error (MAE) and mean squared error (MSE):
\begin{align}
 \text{MAE} = \frac{1}{NH}\sum_{n=1}^N \sum_{h=1}^H \| \bsY_{T+ s(n)+h} - \what{\bsY}_{T+s(n)+h}\|_1 \\
 \text{MSE} = \frac{1}{NH}\sum_{n=1}^N \sum_{h=1}^H \| \bsY_{T+ s(n)+h} - \what{\bsY}_{T+s(n)+h}\|_2^2 
\end{align}
 where $s(n)$ returns the sampled $N$ time index shifts.
We note that the MAE is a suitable evaluation criterion for median forecasts, while the MSE is suitable for mean forecasts, see for more details \cite{gneiting2011making}. As we are predicting means, the MSE should be the preferred evaluation measure. However, the MAE is often regarded as a robust alternative. Thus we consider it as well.

\section{Results}

\begin{figure}
\begin{subfigure}[c]{0.32\textwidth}
\includegraphics[width=.99\textwidth]{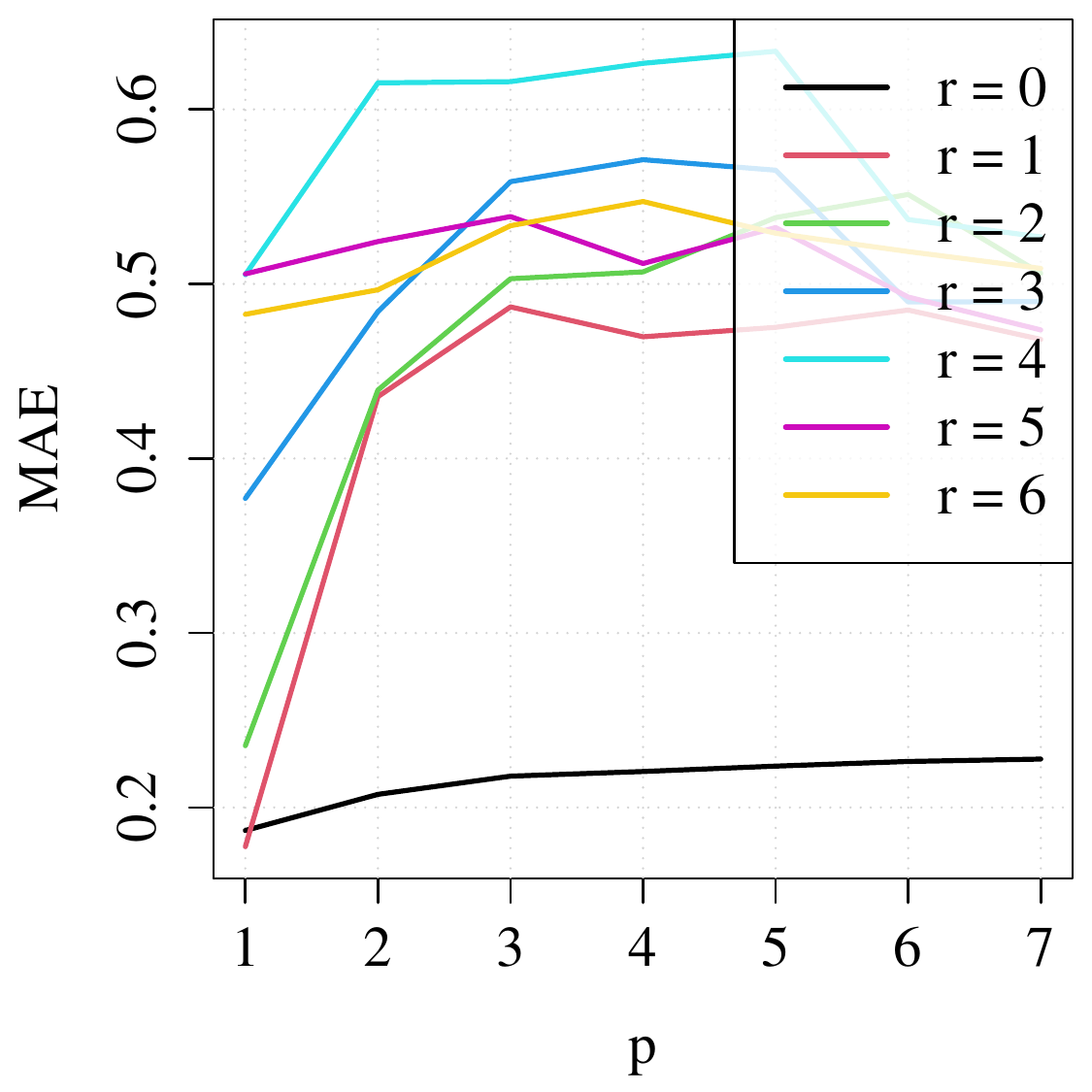} \subcaption{MAE for $T=96$}
\end{subfigure}
\begin{subfigure}[c]{0.32\textwidth}
\includegraphics[width=.99\textwidth]{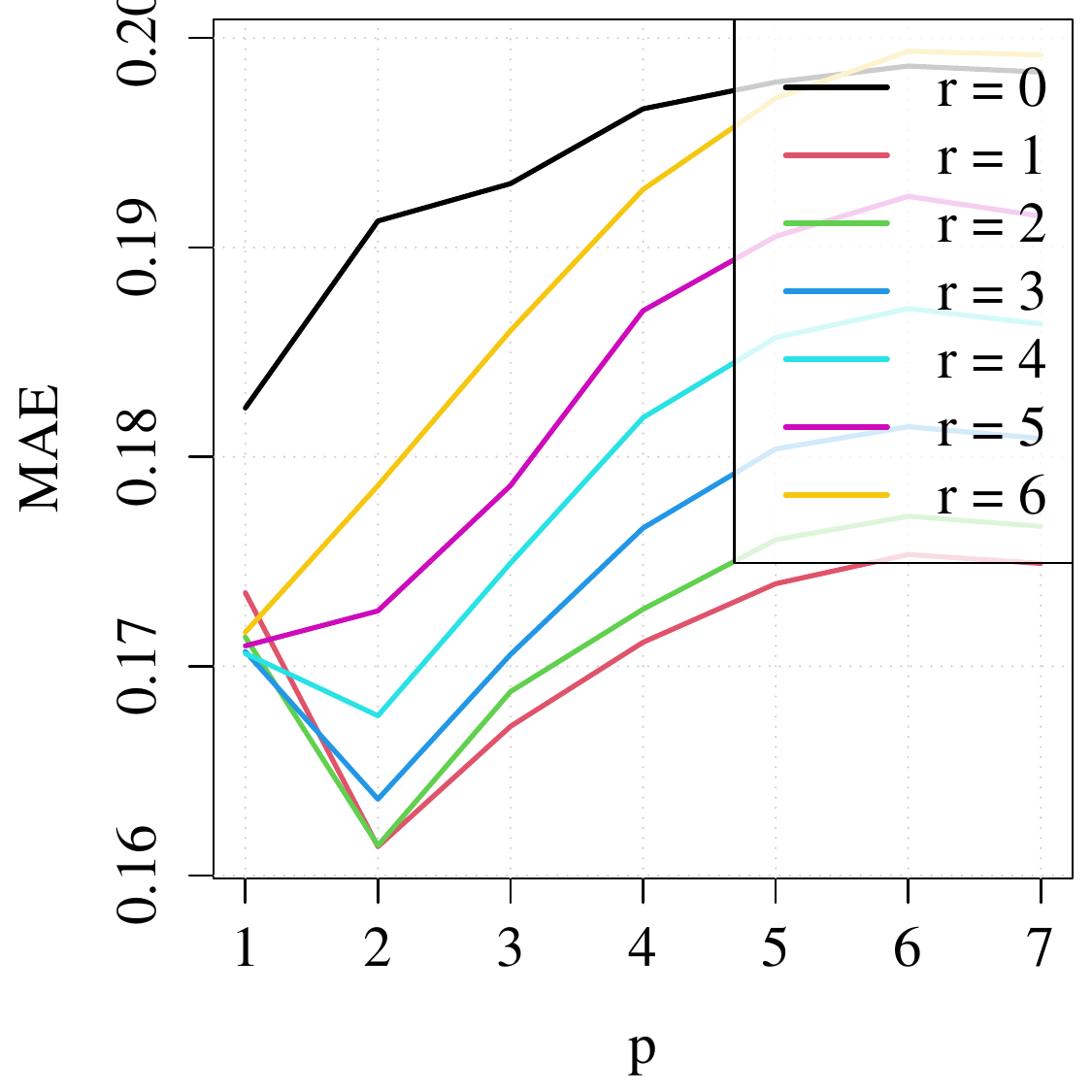} \subcaption{MAE for $T=2\times96=192$}
\end{subfigure}
\begin{subfigure}[c]{0.32\textwidth}
\includegraphics[width=.99\textwidth]{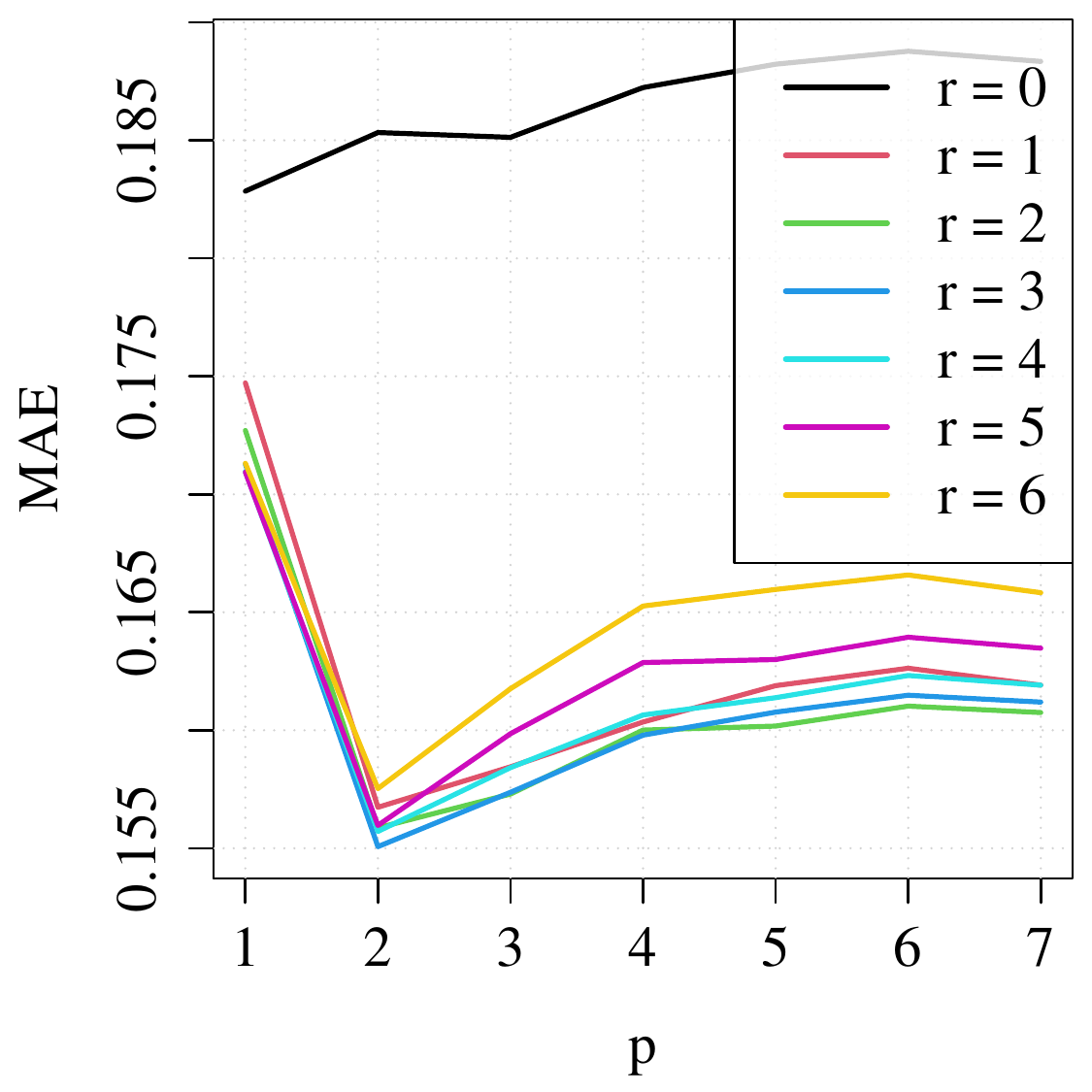} \subcaption{MAE for $T=4\times96=384$}
\end{subfigure}
\begin{subfigure}[c]{0.32\textwidth}
\includegraphics[width=.99\textwidth]{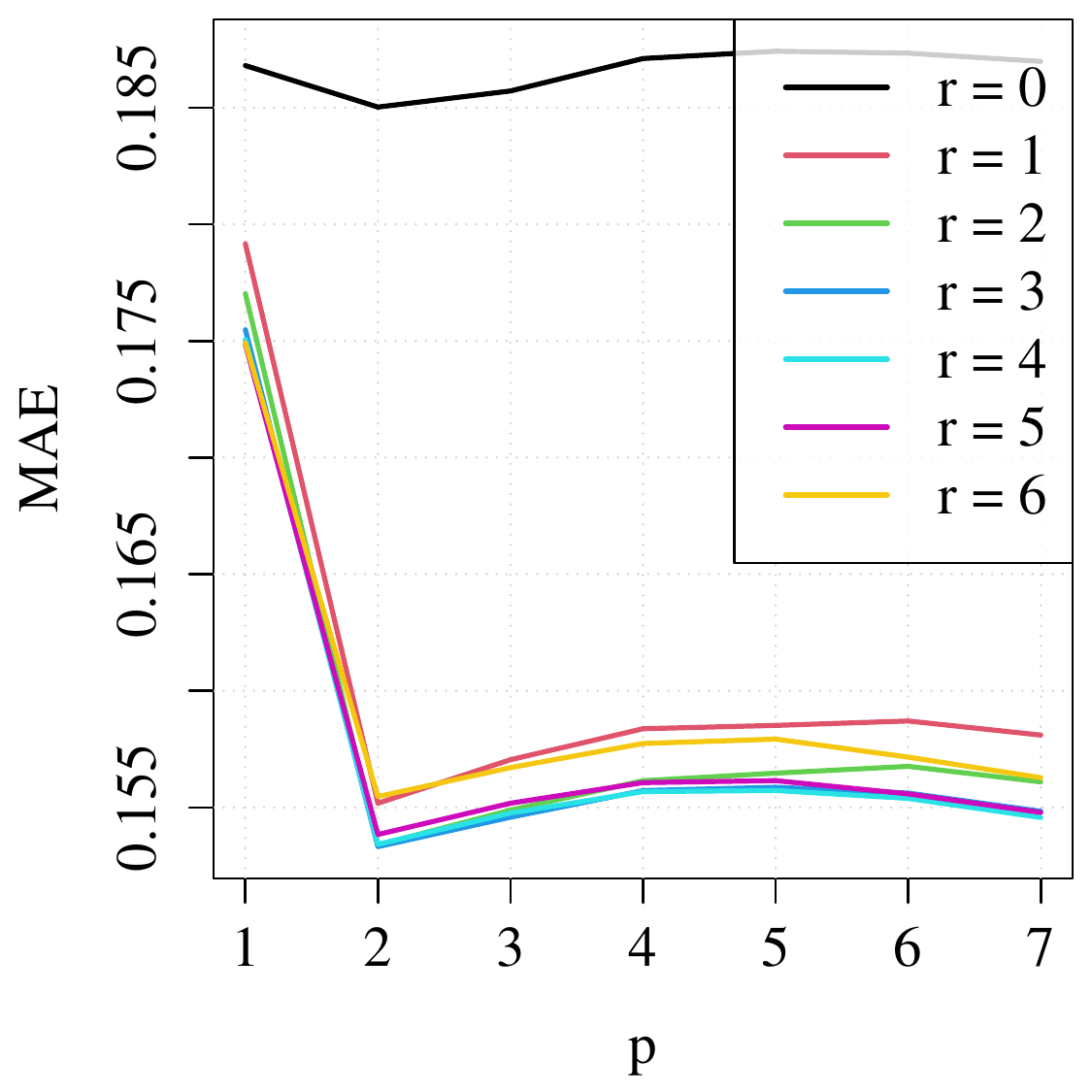} \subcaption{MAE for $T=8\times96=768$}
\end{subfigure}
\begin{subfigure}[c]{0.32\textwidth}
\includegraphics[width=.99\textwidth]{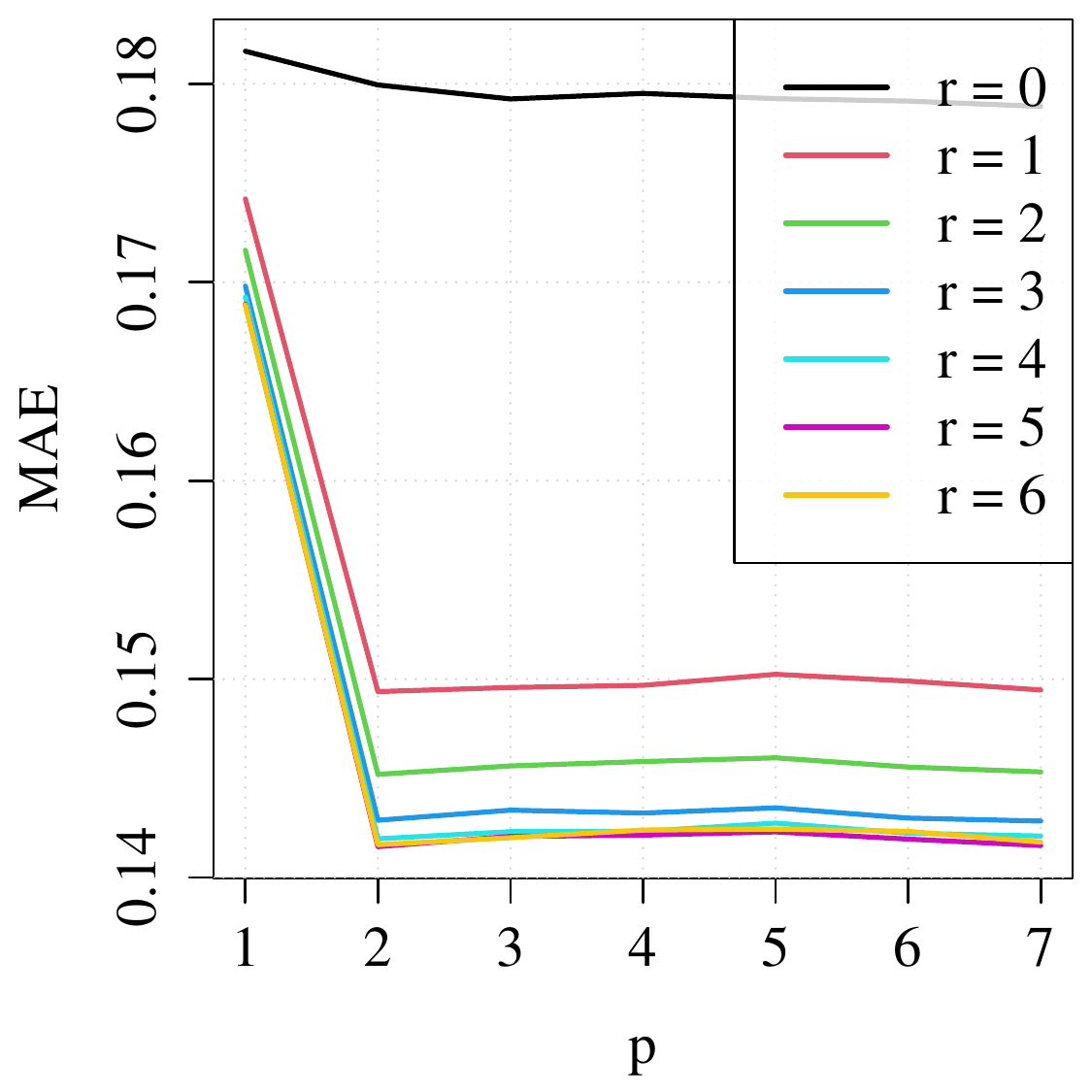} \subcaption{MAE for $T=16\times96=1536$}
\end{subfigure}
\begin{subfigure}[c]{0.32\textwidth}
\includegraphics[width=.99\textwidth]{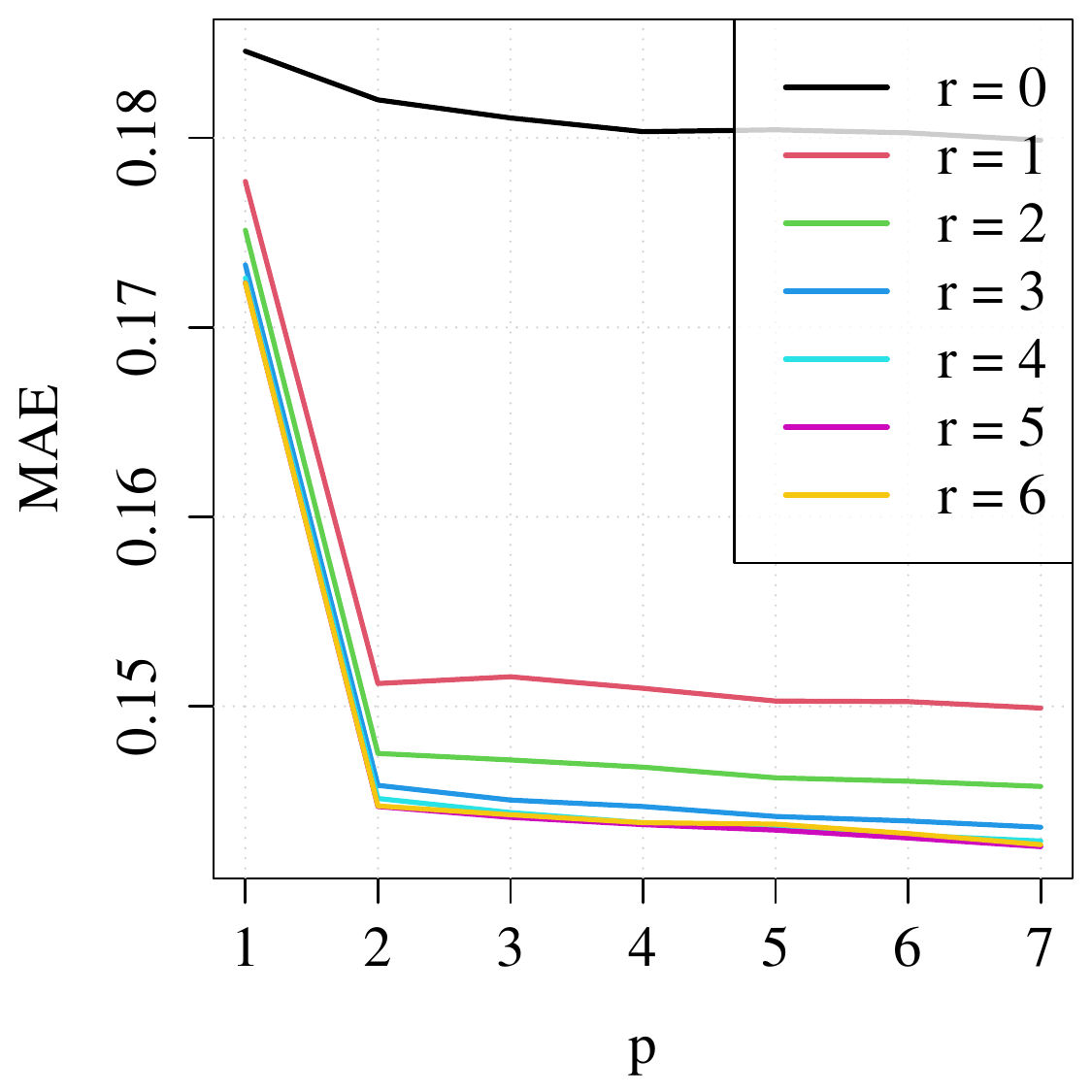} \subcaption{MAE for $T=32\times96=3072$}
\end{subfigure}

        \caption{MAE results for different autoregressive order $p$, cointegration rank $r$ and calibration window length $T$.}
        \label{fig_MAE}
\end{figure}

\begin{figure}
\begin{subfigure}[c]{0.32\textwidth}
\includegraphics[width=.99\textwidth]{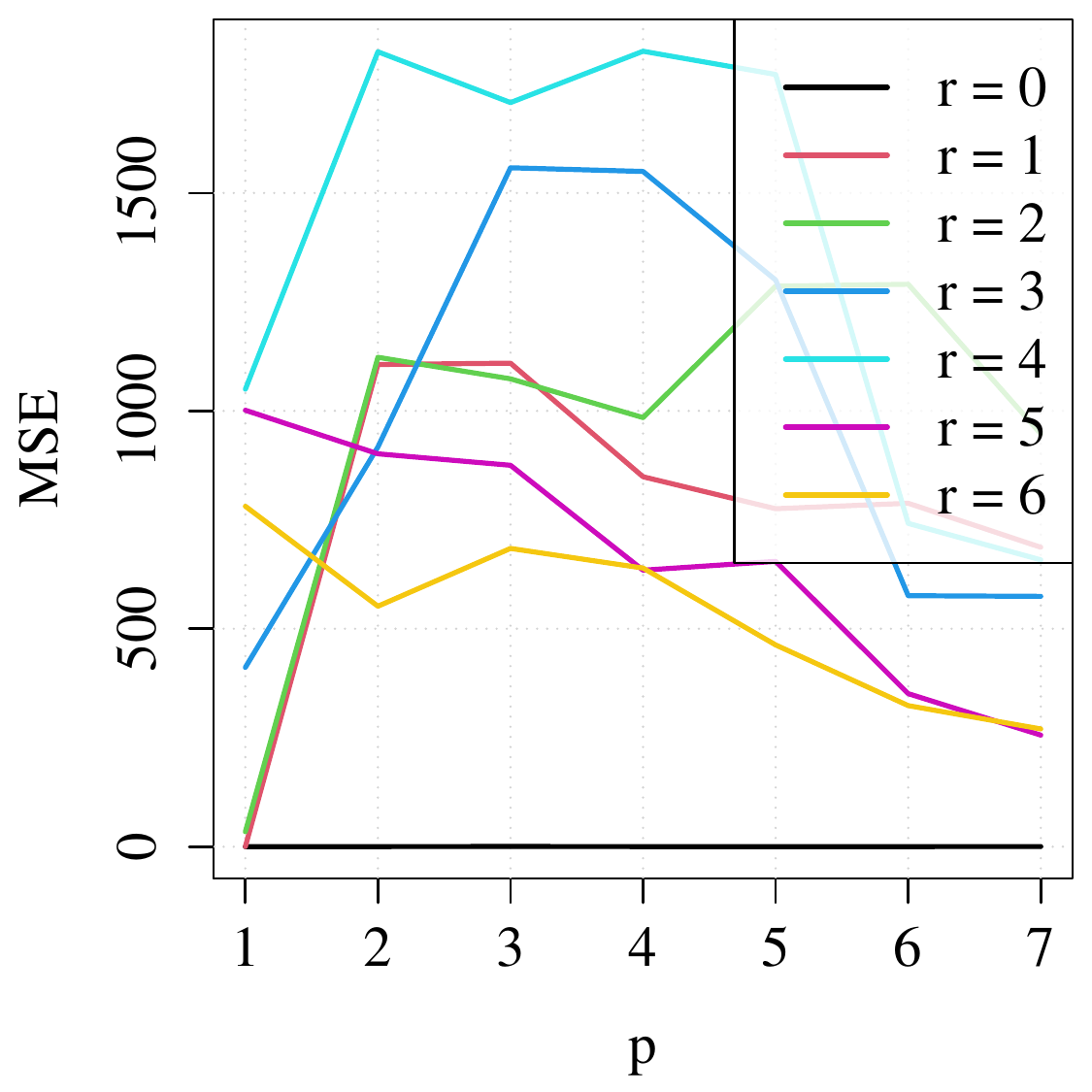} \subcaption{MSE for $T=96$}
\end{subfigure}
\begin{subfigure}[c]{0.32\textwidth}
\includegraphics[width=.99\textwidth]{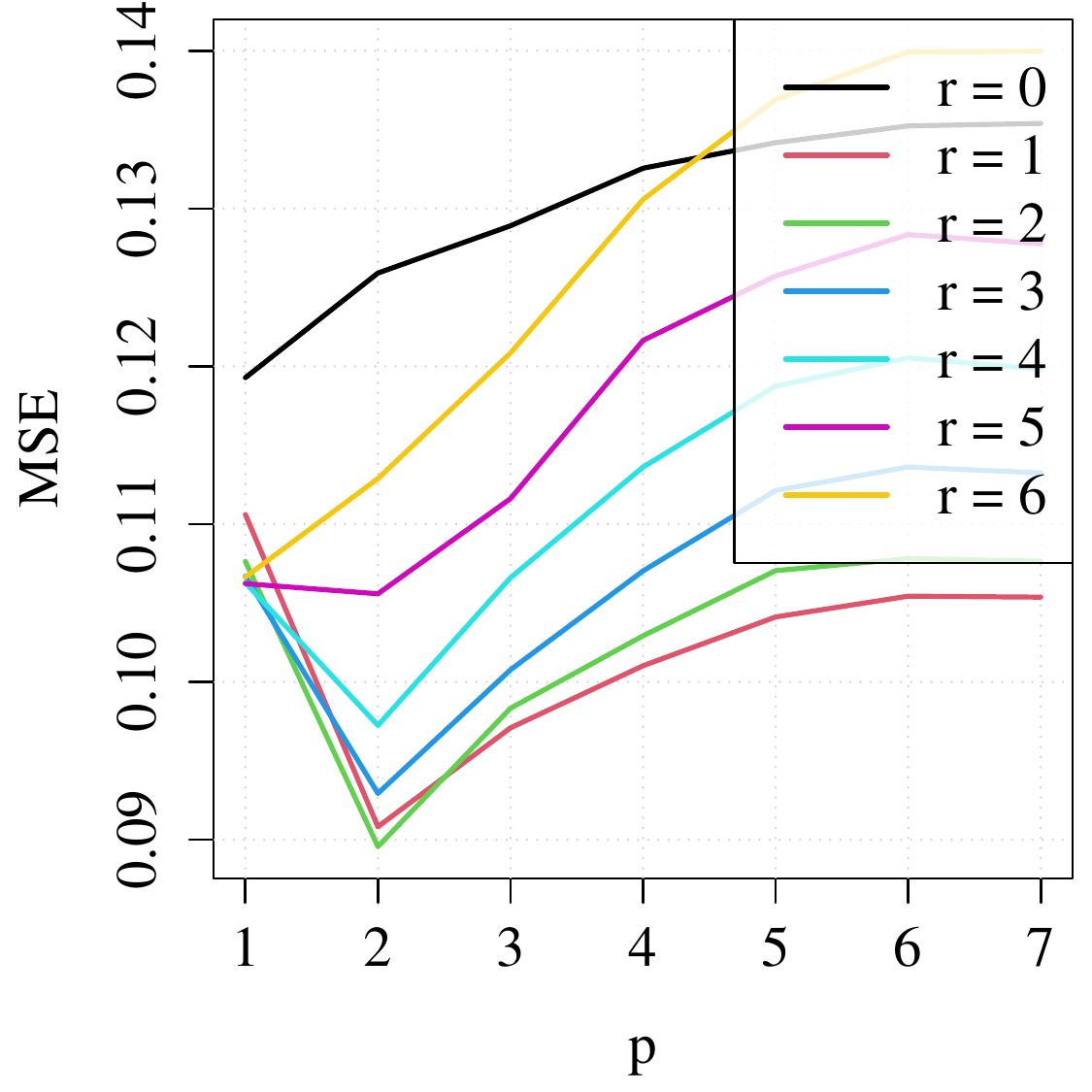} \subcaption{MSE for $T=2\times96=192$}
\end{subfigure}
\begin{subfigure}[c]{0.32\textwidth}
\includegraphics[width=.99\textwidth]{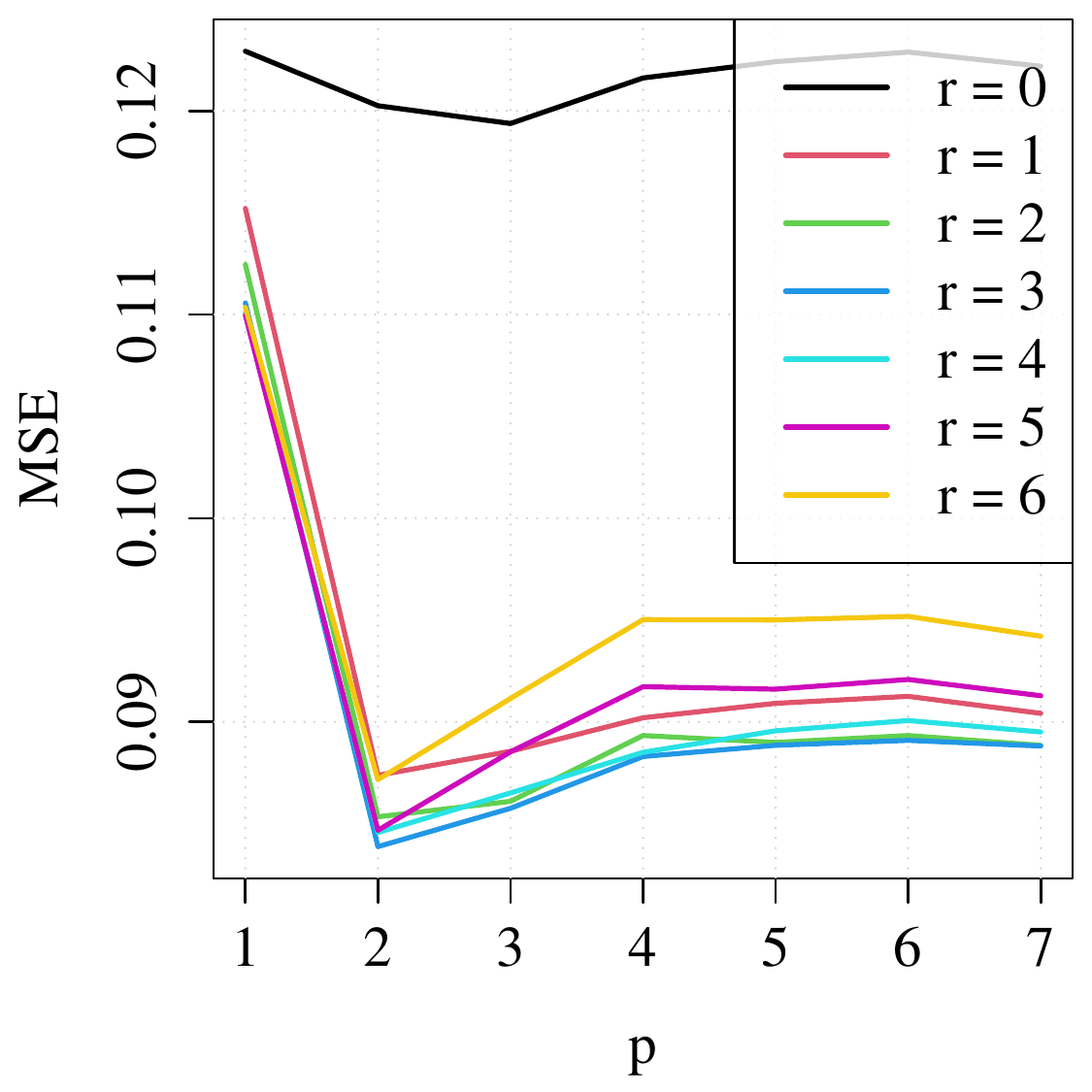} \subcaption{MSE for $T=4\times96=384$}
\end{subfigure}
\begin{subfigure}[c]{0.32\textwidth}
\includegraphics[width=.99\textwidth]{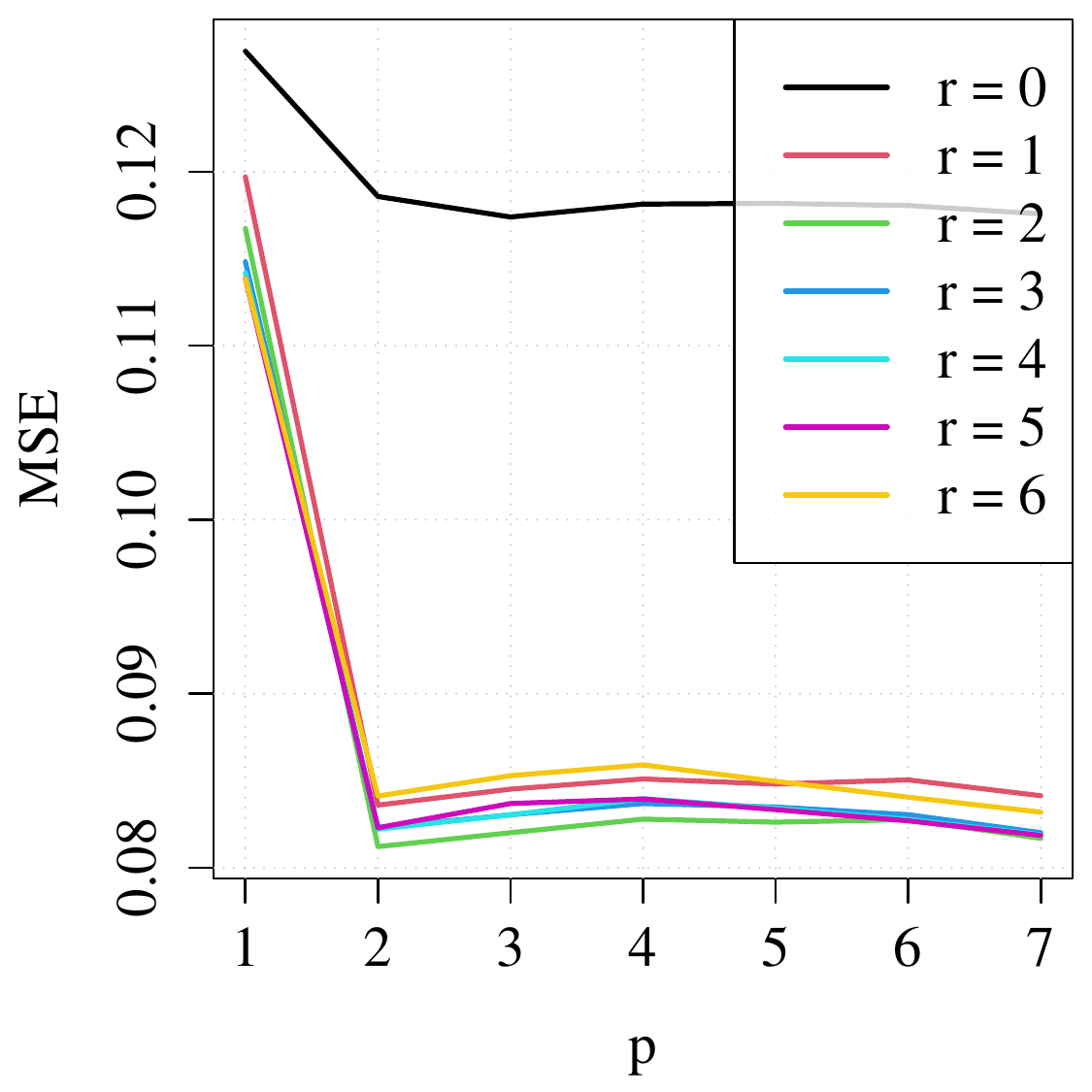} \subcaption{MSE for $T=8\times96=768$}
\end{subfigure}
\begin{subfigure}[c]{0.32\textwidth}
\includegraphics[width=.99\textwidth]{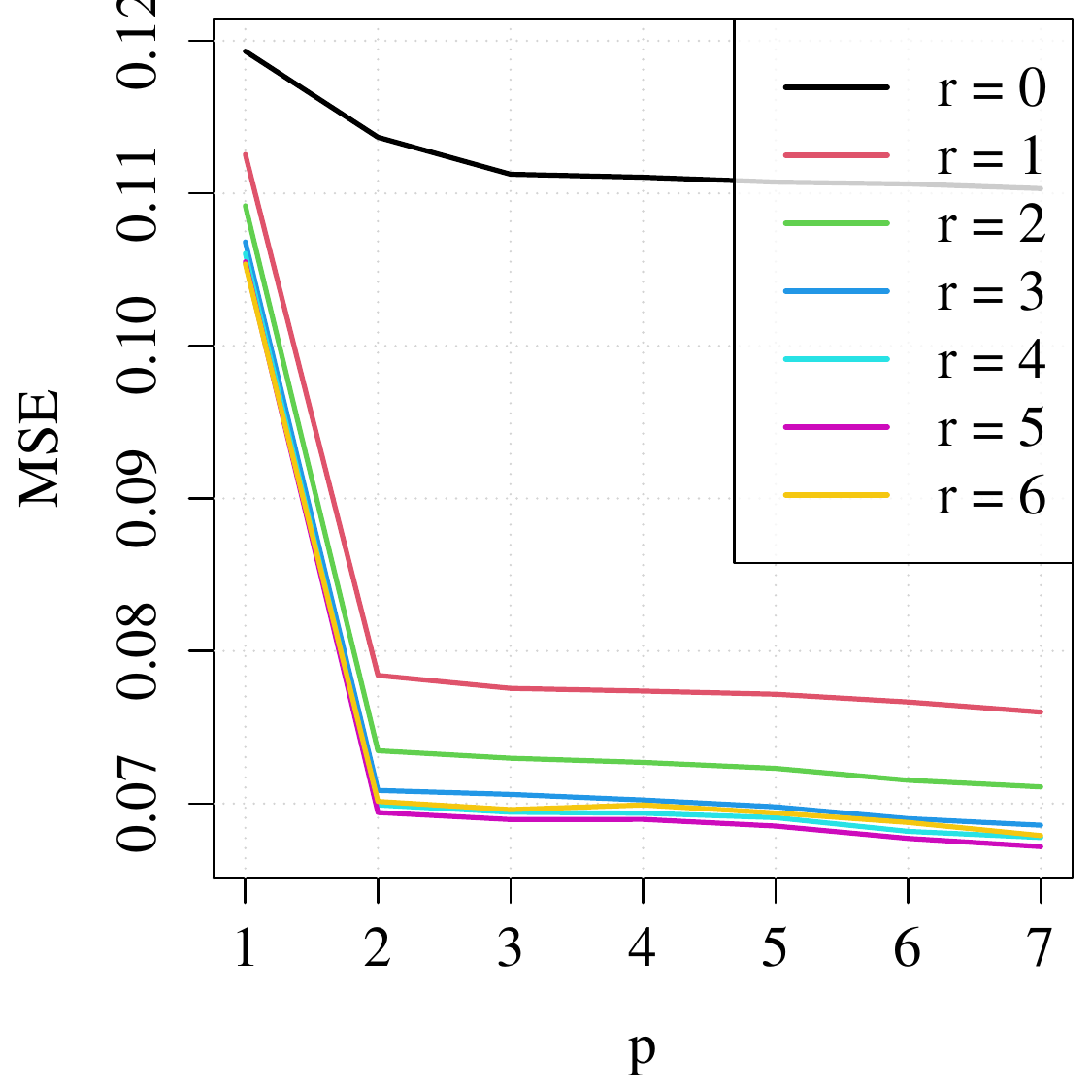} \subcaption{MSE for $T=16\times96=1536$}
\end{subfigure}
\begin{subfigure}[c]{0.32\textwidth}
\includegraphics[width=.99\textwidth]{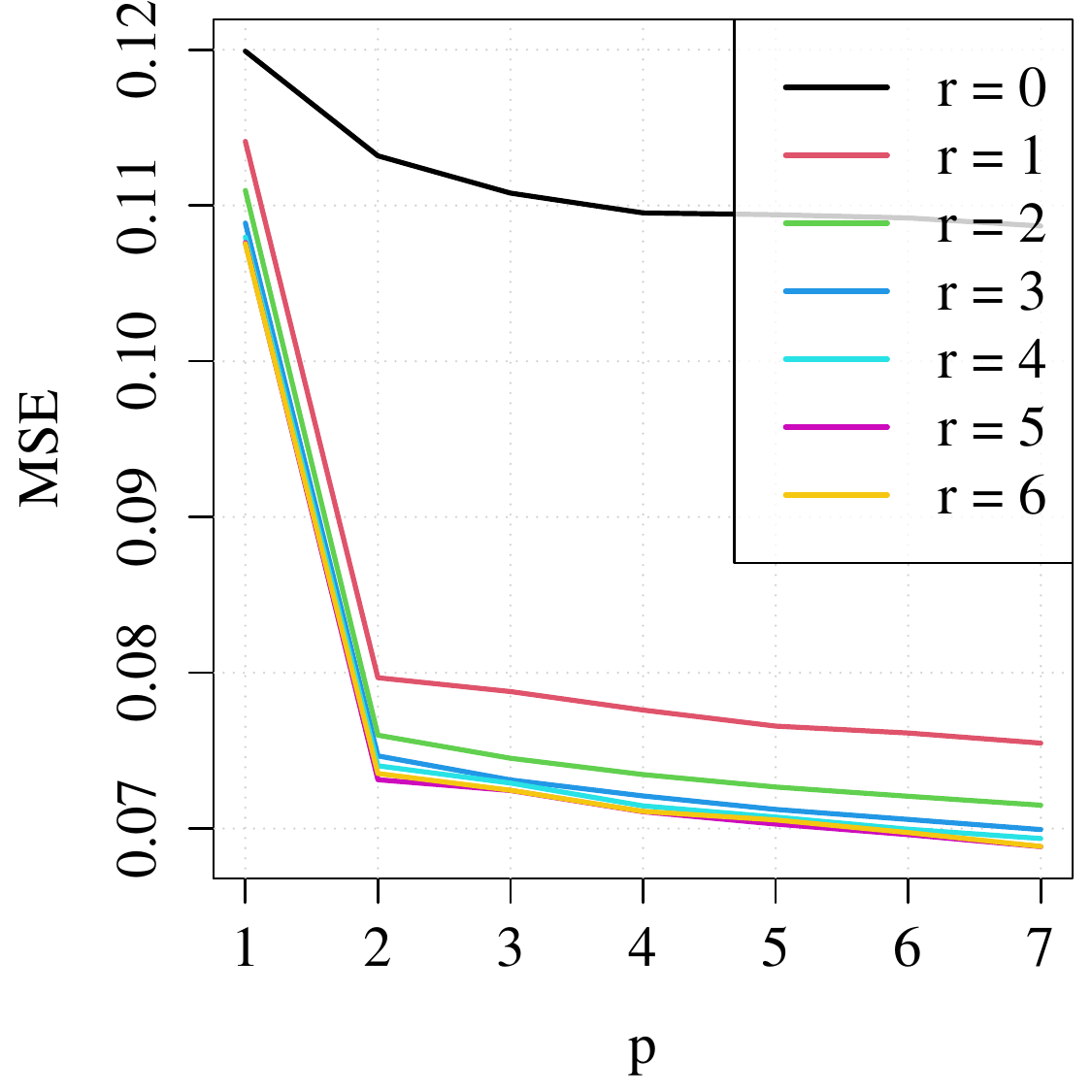} \subcaption{MSE for $T=32\times96=3072$}
\end{subfigure}
        \caption{MSE results for different autoregressive order $p$, cointegration rank $r$ and calibration window length $T$.}
        \label{fig_MSE}
\end{figure}

Figures \ref{fig_MAE} and  \ref{fig_MSE} visualize the MAE and MSE results for different cointegrating ranks $r$, autoregressive orders $p$ and calibration window length $T$.
In addition, Tables  
\ref{tab_mae} and \ref{tab_mse} report the best performing VECM model for a calibration window of size $T$. It also shows the potential improvement with respect to the best performing VAR on $\bsY_t$ and $\Delta \bsY_t$, which correspond to minimal and maximal cointegrating rank, respectively.

In Figures \ref{fig_MAE} and  \ref{fig_MSE} we observe that 
the predictive performance depends substantially on the choice of the calibration window length $T$. The overall pattern for MAE and MSE results are mainly comparable. However, for $T=96$ and MSE the results are suspicious, which is likely due to numerical instability. It can be noted that the minimal cointegrated VECM ($r=0$), i.e. the VAR$(p-1)$ model in first differences, performs poorly for larger $T$. Moreover, we see that the smaller the calibration window length, the better models with low cointegrating rank perform. Thus, cointegration can clearly improve predictive accuracy, best visible for $T=192$ in Fig. \ref{fig_MAE}. For larger calibration window sizes $T$ we also observe that larger autoregressive orders $p$ might be beneficial, which is not surprising as the autoregressive structure can be better learned. Both observations concerning $p$ and $r$ are presented in Tables \ref{tab_mae} and \ref{tab_mse} as well. 

From Figure \ref{fig_MAE} we see clearly a  local optimum for $p=2$ for $T=192$, $T=384$ and $T=768$. This is attained for moderate cointegrating ranks $r=1$ to $r=4$. For larger $T$ we observe that large autoregressive orders $p$ with maximum cointegrating rank $r=d=6$ deliver best predictive performance.
Hence, both model structures: a VECM with rather large $p$ and maximal $r$ (= a VAR on $\bsY_t$), and a VECM with rather small $p$ and $r$ might lead to gain in forecast combination methods.
Indeed, if we combine equally weighted a VECM with $p=7$ and $r=d=6$, and a VECM with $r=1$ and $p=2$ we receive a 1.9\% MAE reduction for $T=768$ (statistically significant with respect to the Diebold-Mariano test). The reason is likely that both models capture different model characteristics adequately. The VAR model the stationary behavior and the VECM with small $r$ the relevant instationarity structure.


The reason why cointegrated models perform well can be inferred
from Tables \ref{tab_mae} and \ref{tab_mse} as well. 
For small $T$ we see large improvement against the best VAR model for  $\bsY_t$. 
Very competitive VAR models on $\bsY_t$ ($r=d=6$) require large calibration window length $T$ to ``learn'' the stationarity behaviour adequately. In contrast, the assumption of a unit root in every component ($r=0$) adds additional information to the data which reduces the estimation risk in the corresponding VAR models on $\Delta \bsY_t$. This is very beneficial in terms of predictive accuracy for small $T$. However, there are calibration window sizes where the optimal MAE and MSE is not attained for limiting cases ($r=d=6$ or $r=0$). For $T=192$ we observe the largest possible improvements due to the incorporation of cointegration. The MAE diminishes by 5\% and MSE by 15\% compared to the best alternative. This is very encouraging for future model building.

\begin{table}[ht]
\centering
\begin{tabular}{r|rrrrrr}
T/96 (=length in days) & 1 & 2 & 4 & 8 & 16 & 32 \\ 
  \hline
  Best $p$ & 1 & 2 & 2 & 2 & 2 & 7 \\ 
  Best $r$ & 2 & 2 & 4 & 4 & 6 & 6 \\ 
  Improvement to best VAR on $\Delta \bsY_t$ & 0.00 & 0.05 & 0.09 & 0.12 & 0.16 & 0.17 \\ 
  Improvement to best VAR on $ \bsY_t$ & 0.22 & 0.08 & 0.04 & 0.01 & 0.00 & 0.00 \\ 
   \hline
\end{tabular}
\caption{Summary MAE results of the forecasting study, for different calibration window length.}
\label{tab_mae}
\end{table}

\begin{table}[ht]
\centering
\begin{tabular}{r|rrrrrr}
T/96 (=length in days) & 1 & 2 & 4 & 8 & 16 & 32 \\ 
  \hline
  Best $p$ & 1 & 2 & 2 & 2 & 7 & 7 \\ 
  Best $r$ & 2 & 3 & 4 & 3 & 6 & 6 \\ 
  Improvement to best VAR on $\Delta \bsY_t$ & 0.00 & 0.16 & 0.24 & 0.29 & 0.36 & 0.36 \\ 
  Improvement to best VAR on $ \bsY_t$ & 0.62 & 0.15 & 0.06 & 0.01 & 0.00 & 0.00 \\ 
   \hline
\end{tabular}
\caption{Summary MSE results of the forecasting study, for different calibration window length.}
\label{tab_mse}
\end{table}

\section{Summary and outlook}

We have shown that incorporation of cointegration effects in VAR-type models for wind power forecasting can improve forecasting accuracy significantly. Especially for the short calibration windows the  improvements can be substantial. In the empirical study on German quarter-hourly wind data the considered VECM reduce the MAE of the forecast by 5\% and MSE by 15\% in comparison to the best VAR-type models trained on the original data in levels and on the first-differenced data, respectively. Moreover, there is some evidence that model combination may improve the results further when combining stationary VAR-type models with large autoregessive order with non-stationary VECMs with low autoregressive order and small cointegrating rank. What needs to be considered as well is formal determination of the cointegrating rank and possibly allowing it to be time-varying as in \cite{arce2019}.
Future research in this area will also focus on investigating and utilizing the long-memory properties of wind speed and wind power data, potentially leading us to employ fractionally cointegrated VAR models.

\bibliographystyle{apalike}
\bibliography{bib}

%

\end{document}